\documentclass[epjCONF, onecolumn]{svjour}
\usepackage{graphics}
\usepackage[varg]{txfonts} 
\usepackage[latin1]{inputenc}
\session-title{Assembling the Puzzle of the Milky Way}
\begin{document}
\title{The puzzling assembly of the Milky Way halo -- contributions from dwarf Spheroidals  and globular clusters}
\author{Andreas Koch\inst{1}\fnmsep\thanks{\email{akoch@lsw.uni-heidelberg.de}} \and 
Sebastien L\'epine\inst{2}\and
\c Seyma \c{C}al{\i}\c{s}kan\inst{1,3} 
 }
\institute{Zentrum f\"ur Astronomie der Universit\"at Heidelberg, Landessternwarte, K\"onigstuhl 12, 69117 Heidelberg, Germany
\and Department of Astrophysics, Division of Physical
  Sciences, American Museum of Natural History, Central Park West at
  79th Street, New York, NY 10024, USA
\and Department of Astronomy and Space Sciences, Ankara University, 06100, Tando\u gan, Ankara, Turkey
}
\abstract{
While recent sky surveys have uncovered large numbers of ever fainter 
Milky Way satellites, their classification as star clusters, low-luminosity
galaxies, or tidal overdensities remains often unclear. Likewise, their 
contributions to the build-up of the halo is yet debated. 
In this contribution we will discuss the current knowledge of the stellar populations 
and chemo-dynamics in these puzzling satellites, with a particular focus on 
dwarf spheroidal galaxies and the 
globular clusters in the outer Galactic halo. Also the question of whether some of the outermost  
halo objects are dynamically associated with the (Milky Way) halo at all 
is addressed in terms of proper measurements in the remote Leo I and II dwarf galaxies. 
} 
\maketitle
\section{Introduction}\label{intro}
Searle \& Zinn's (1978) picture of a hierarchical assembly of galaxies like the Milky Way (MW) has been bolstered by the discoveries of large numbers of ever fainter satellites around the MW and M31 
in  recent,  ambitious sky surveys. 
These systems range from relatively luminous  dwarf spheroidal (dSph) galaxies towards ever fainter objects, commonly dubbed {\em ultrafaint} dwarfs (UFDs; 
e.g., Zucker et al. 2006; Belokurov et al. 2007; Walsh et al. 2007;  Majewski et al. 2007; McConnachie et al. 2008; see also Koch 2009, and references therein). 
At $10^3$--$10^5$ M$_{\odot}$, the stellar masses of the UFDs are comparable to the most extended MW star clusters. 
Intriguingly, those globular clusters (GCs) with the largest radii, in the transition regime between GCs and UFDs (e.g., Fig.~1 in Misgeld \& Hilker 2011),  
 are predominantly found in the {\em outermost} MW halo\footnote{Those clusters are also typically younger than inner halo clusters with otherwise comparable properties.}. 
In fact, current scenarios envision a dichotomy of an  inner halo, formed {\em in situ}, and an outer, accreted component. 
In the following we will  tackle  the ``puzzle''  of the Galactic halos --  the formation history of the entirety (read: the  MW halo) -- 
by studying its complexity of constituents (i.e., its halo GCs and dSph satellites). 
In particular, we address the discrimination between UFDs and the extended outer halo GCs and  their role in assembling the Galactic halo. 
\section{Winnowing dSphs from star clusters}\label{sec:1}
DSph galaxies have always been characterized as low-luminosity systems (see reviews by Koch 2009; Tolstoy et al. 2009). 
Some noteworthy key features of the dSphs are, amongst others, their low luminosities, their high dark matter content (with mass-to-light ratios, $M/L$, of up to several thousands), 
the omnipresence of old ($>$12 Gyr) 
stellar populations, 
low metallicities, a slow chemical evolution, and generally complex star formation histories (Grebel 1997; Tolstoy et al. 2009). 
However, for some objects it is yet unclear whether they are truly old and metal poor systems like the dSphs, or very extended, (tidally) perturbed 
 stellar systems, and thus essentially dying star clusters, free of dark matter, or mere density enhancements in tidal streams. 
In the latter cases, 
 the ``missing satellite problem''  would remain a problem (e.g., Bovill \& Ricotti 2009). 

The dSphs have only experienced slow chemical evolution and little chemical enrichment, rendering them metal poor systems. As Fig.~1 (left panel) shows, they follow a well defined metallicity-luminosity relation 
that extends down to the faintest galaxies 
\begin{figure}
\resizebox{1\columnwidth}{!}{
  \includegraphics{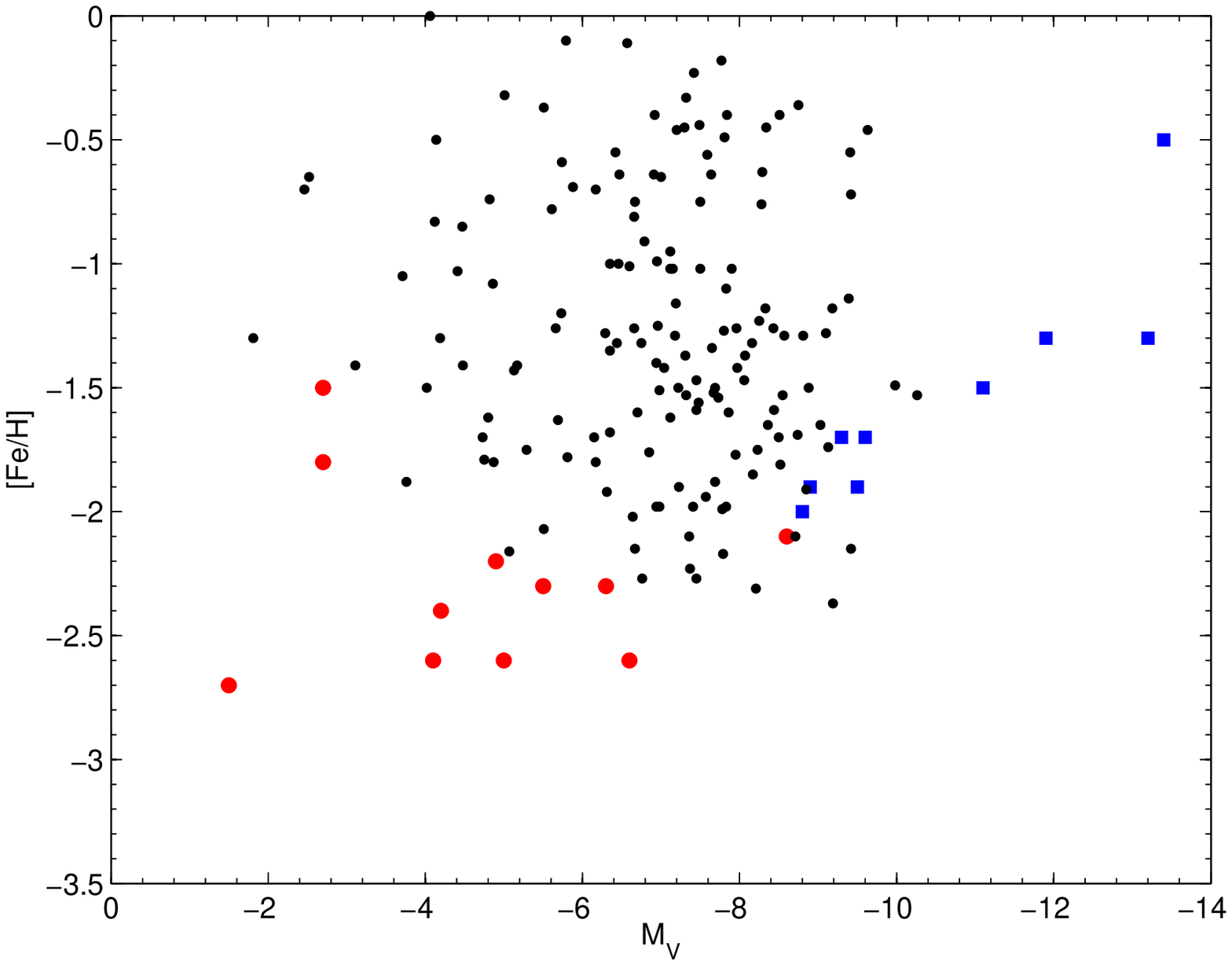} 
  \includegraphics{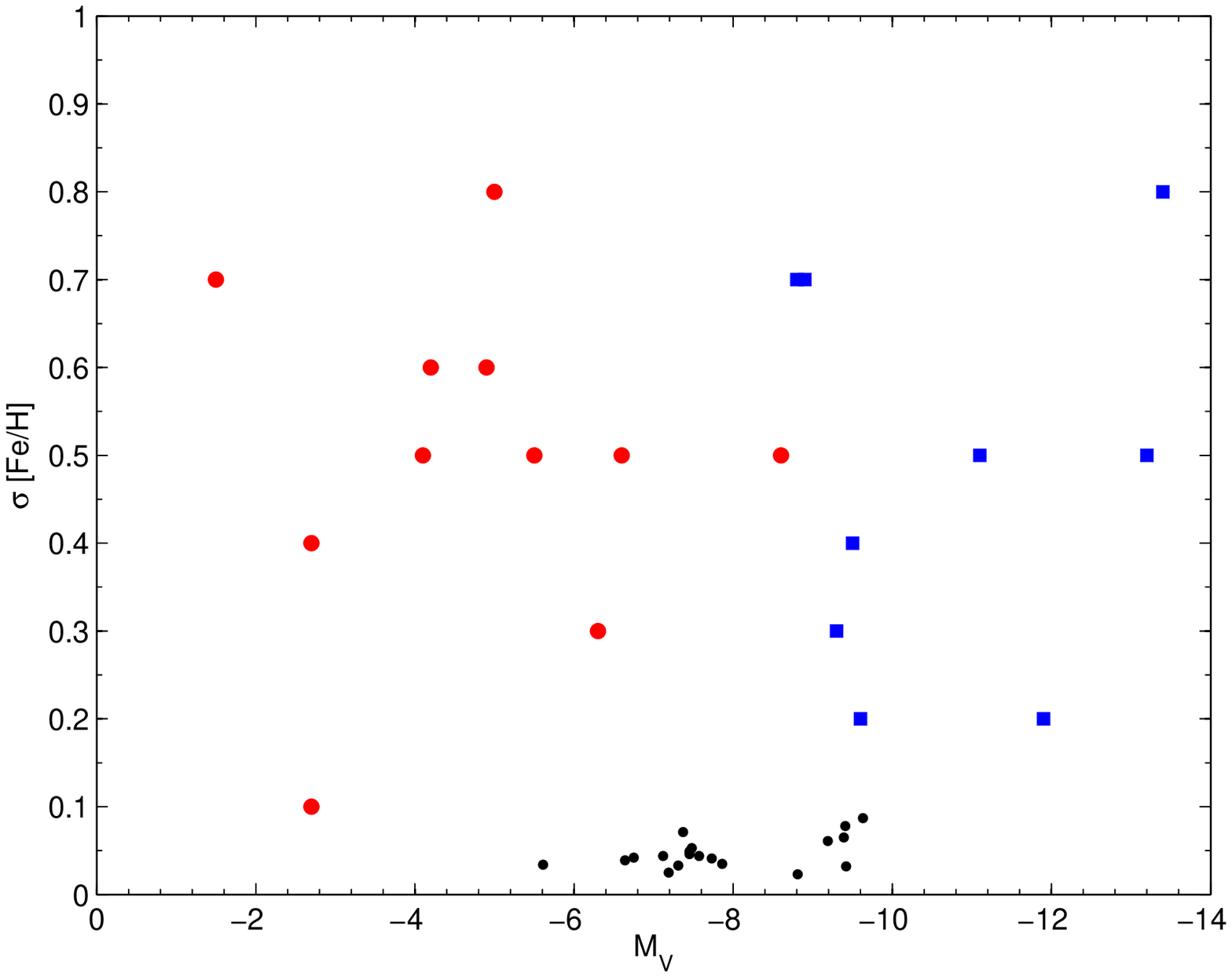} 
  }
\caption{Mean metallicity (left panel) and 1$\sigma$ spreads (right panel) of Galactic GCs (black dots), the classical dSphs (blue symbols), and ultrafaint dSphs (red points); using data from Koch 2009 (and references therein); Harris 1996 [2011 version]; and Carretta et al. (2009).}
\label{fig:1}       
\end{figure}
The simple reason for such a correlation is that the dSphs possess deep (dark matter) potential wells, in which gas can be efficiently 
retained for further enrichment. GCs, on the other hand,  are dark matter free and no such relation exists. They rather cover a broad range of metallicities\footnote{Also note that star clusters do not contain any very metal poor stars below $
\sim -2.4$ dex.}   irrespective of their luminosity: a low metallicity alone does not signify a dSph.   

On the other hand, the deep potentials of the dSphs enable prolonged star formation and enrichment of subsequent generations of stars with the retained metals of the previous generations. As a  consequence, 
 the dSphs exhibit  abundance spreads of several tenths of a dex, which is in clear contrast to the GCs that are, to first order, considered mono-metallic\footnote{Currently, evidence for multiple generations 
of stars in GCs is accumulating and  some light chemical elements are found to vary within any given cluster due to their specific internal evolutionary and enrichment  histories. For instance, 
Carretta et al. (2010) suggested the presence of a Na-O anti-correlation as a defining factor for a ``GC''.} (Fig.~1, right panel). 

We mention here two prime examples, for which a clear-cut classification has been controversial since their discoveries. 
Firstly, it has been suggested that the faint object Segue~1 could be a dissolving star cluster, associated with the Sagittarius (Sgr) dwarf; overlap (on the sky and in radial velocity) would lead to an inflated velocity dispersion so that the inferred high $M/L$ fails as an unambiguous indicator of a dark matter dominated dSph (Niederste-Ostholt et al. 2009). 
Subsequently, Simon et al. (2011) measured $M/L\sim3400$, which is not explicable by contamination with Sgr stars alone. Segue~1  
has a low, mean iron abundance of $-2.7$ dex and a 1$\sigma$ iron spread of 0.7 dex. 
Moreover, the full abundance ranges, e.g., in [C/H] and [Fe/H] are in excess of 1.5 dex, thus 
spanning a factor of several tens in the (heavy) element content (Norris et al. 2010; see also the contribution by G. Gilmore in this Volume),  
pointing to a dark matter dominated system (i.e., the potential well was deep enough to allow for chemical self-enrichment). 

Another example of this class  is Bo\"otes II (Walsh et al. 2007; Koch et al. 2009a): based on low-resolution spectra of 5 member-stars, the latter work finds a mean metallicity and radial velocity dispersion consistent with a dark matter dominated, old, and metal poor dSph-like population; however, Bo\"otes~II lies square on the leading arm of Sgr in projected location on the sky, radial velocity, and distance. 
Every possible chemical abundance information, ideally for large numbers of stars, 
 is thus required to describe the chemical evolution of such objects  to relate them to the earliest galactic enrichment phases and to assess 
 their role to the build-up of the halo.
\section{Outer halo GCs}\label{sec:2}
As elaborated above, ideally, we need to monitor the chemical  abundance patterns and search for spreads in the faint structures to fully characterize their nature. Likewise, the chemical abundance patterns of GCs in the outer MW halo bears vital information 
about the  formation and assembly  history of the Galaxy.  
The need to obtain spectroscopy of single stars in the remote GCs of the outer halo (say, $>$30 kpc) suffers from two main problems: 
At a first glance, these systems appear sparse and they are, on average, spatially more extended than GCs of the inner halo (e.g.,  Martin et al. 2008). 
Na\"ively, this seems ideal to easily obtain uncontaminated spectroscopy for statistically significant samples of stars, preferably in multi-object mode.
As the example of an {\em inner halo} cluster, NGC~6397 (at $R_{\odot}$=2.3 kpc), in Fig.~3 (left panel) shows, this is common practice and these objects are spectroscopically well studied (e.g., Lind et al. 2011)
 -- present-day instruments like the VLT/FLAMES multiobject spectrograph can accommodate more than 100 fibres across a field of view of $\sim$25'.  This allows us to target stars out to the tidal radius without running out of 
sources due to crowding, source confusion, or fibre crossings. 
\begin{figure}
\resizebox{0.95\columnwidth}{!}{
  \includegraphics{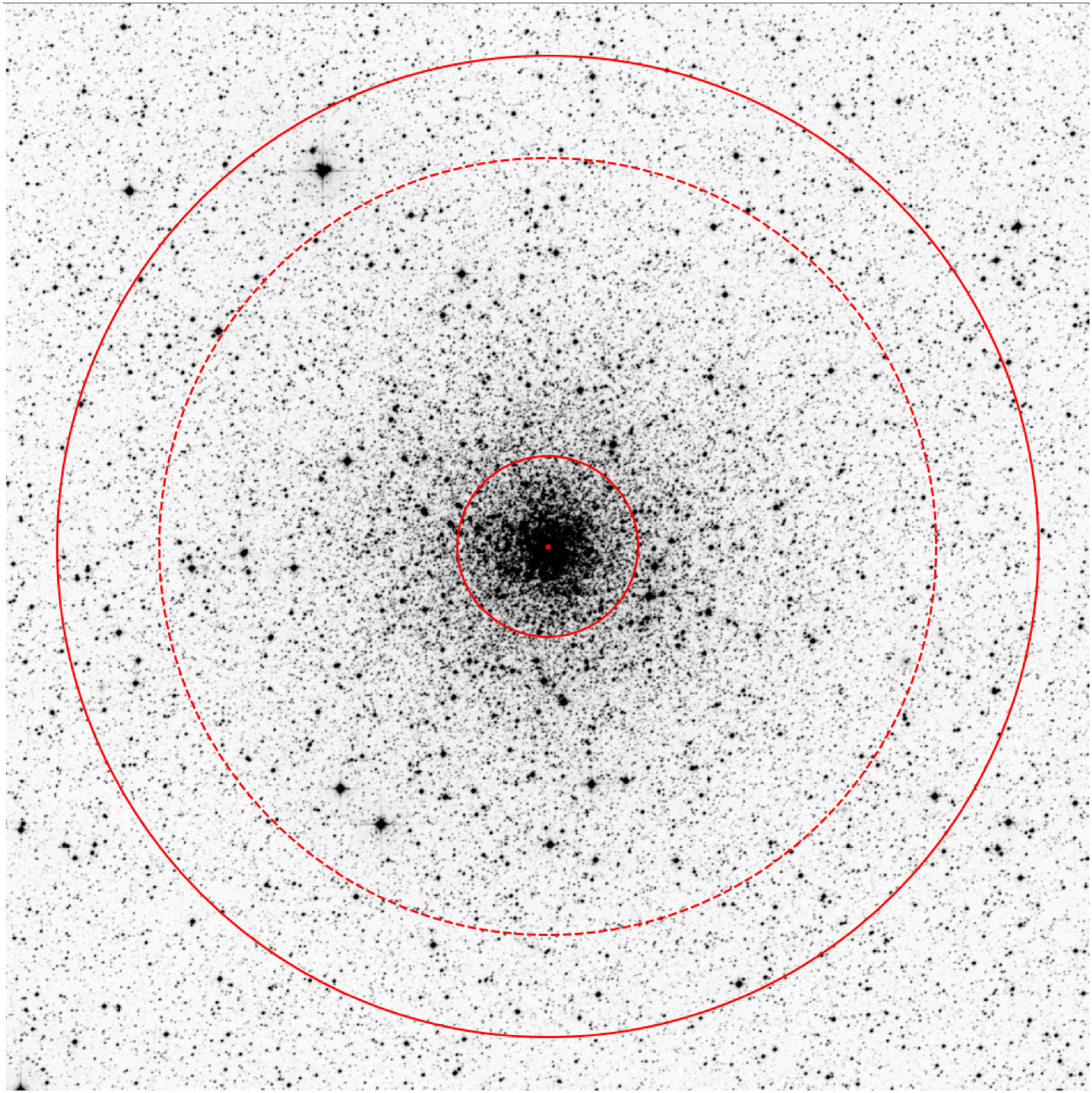} 
  \includegraphics{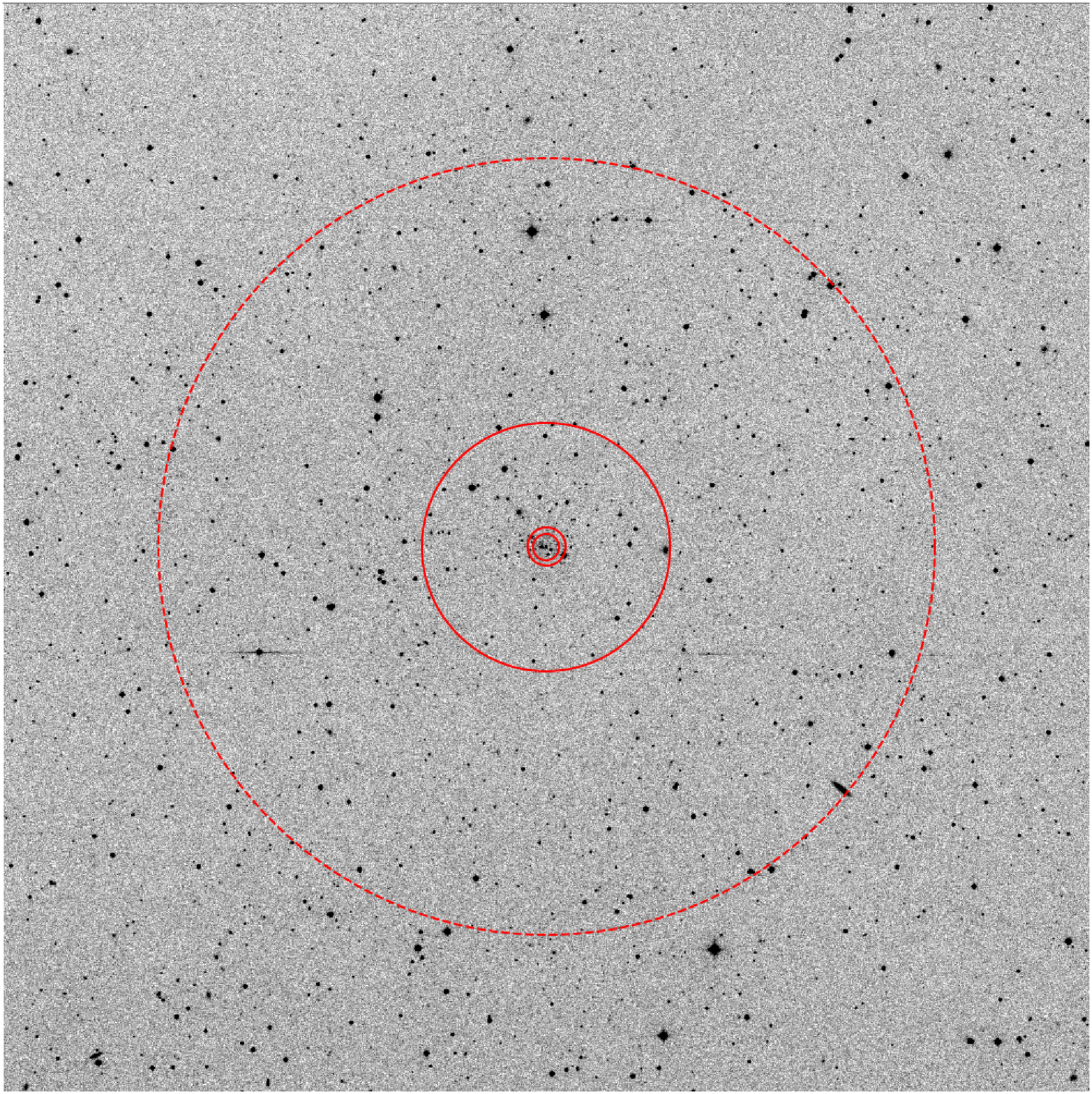} 
  }
\caption[]{DSS images\footnotemark\  of the core-collapsed (r$_c$=0.03 pc; r$_h$=3 pc; r$_t$=16 pc), close (R$_{\odot}$=2.3 kpc) GC NGC 6397 (left panel), and the remote (R$_{\odot}$=93 kpc), extended (r$_c$=11 pc; r$_h$=16 pc; r$_t$=108 pc) cluster Pal 3 (right panel). Solid lines indicate the aforementioned radii, while the dashed circle shows the field of view of the FLAMES spectrograph (25' diameter). Each image extends $17'\times17'$.}
\label{fig:2}    
\end{figure}
\footnotetext{{\em Copyright Note:} Based on photographic data obtained using The UK Schmidt Telescope. The UK Schmidt Telescope was operated by the Royal Observatory Edinburgh, with funding from the UK Science and Engineering Research Council, until 1988 June, and thereafter by the Anglo-Australian Observatory (AAO). Original plate material is copyright \copyright~of the Royal Observatory Edinburgh and the AAO. The plates were processed into the present compressed digital form with their permission. The Digitized Sky Survey was produced at the Space Telescope Science Institute under US Government grant NAG W-2166.
Plates from this survey have been digitized and compressed by the STScI. 
The digitized images are copyright  \copyright~1993-2000 by the
AAO Board, and are distributed herein by
agreement. All Rights Reserved. All material not subject to the above copyright provision is
copyright \copyright ~2000 by the Association of Universities for Research in Astronomy, Inc.  All Rights Reserved.
Produced under Contract No. NAS5-2555 with the National Aeronautics 
and Space Administration.
}
However, the latter becomes problematic for the remote systems -- one has to bear in mind that even the remarkably large radial extent of these GCs translates into 
a mere few minutes of arc on the sky, given their large distances. In the case of the outer halo GC Pal~3 ($R_{\odot}$=93 kpc), half-light and tidal radii correspond to 0.4' and 4', respectively (Hilker 2006), 
rendering multi-object approaches an inefficient strategy. 

Secondly, the (often sparse) red giant branches of the remote GCs are faint and high-resolution spectroscopy requires long exposure times at $\ge$6-m class telescopes (Koch et al. 2009b\footnote{This work contained individual spectra down to V=18.5 mag, for which integration times of $\sim$4 hours were required to reach moderate S/N ratios of at least 30 for a handful of stars.}; Cohen et al. 2011).  
An alternative path to determining chemical abundance ratios for the faint residents of the outer halo  is then to obtain integrated cluster spectra (McWilliam \& Bernstein 2008) or to 
co-add low S/N spectra of many stars to extract a {\em mean} abundance information (Koch et al. 2009b; Koch \& C\^ot\'e 2010). The abundance spread can then be evaluated  in a statistical manner.  
In Fig.~3 we highlight the current knowledge of the chemical inventory of the outer halo GCs in comparison with halo field stars and some archetypical inner halo GCs (R$_{\rm GC}\lesssim$ 12 kpc). This is done for the 
[$\alpha$/Fe] and [Y/Ba] abundance ratios -- important tracers of the chemical evolution of any stellar system (e.g., Tolstoy et al. 2009).
\begin{figure}
\begin{center}
\resizebox{0.8\columnwidth}{!}{
  \includegraphics{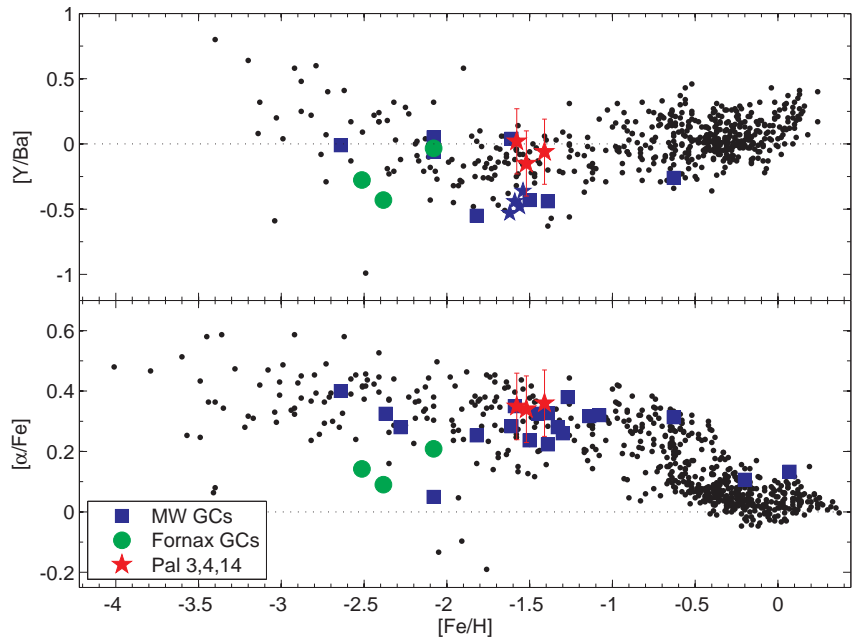} 
  }
\end{center}
\caption{Comparison of the [$\alpha$/Fe] abundance ratios in Galactic field stars (black dots), the Fornax dSph (Letarte et al. 2006), and MW GCs, with a particular focus on the outer halo clusters Pal 3, 4, and 14 (at R$_{\rm GC}$=96, 111, and 72 kpc). 
See Koch et al. 2009b, 2010, and \c{C}al{\i}\c{s}kan et al. (2011) for details and references.}
\label{fig:3}      
\end{figure}
As already discussed in Koch et al. (2009b) and Koch \& C\^ot\'e (2010), 
these systems do not stand out in any (chemical) regard, safe for a dominance of r-process nucleosynthesis (Koch et al. 2009b; \c{C}al{\i}\c{s}kan et al. 2011). 
For instance, the [$\alpha$/Fe] and other heavy element abundance patterns are fully compatible with those in the inner halo, indicating that the  inner and outer MW halos have evolved, 
at least chemically, homogeneously -- which contrasts their otherwise diverse   
characteristics such as a younger age and larger spatial extent compared to the inner systems. 
This is not self evident as individual systems at large distances show anomalies that question their status as genuine halo GCs and rather suggests an accretion origin (Lee et al. 2005; Cohen et al. 2011).   
While the occurrence of an abundance spread cannot be excluded for at least one remote, extended GC, NGC 2419 (Cohen et al. 2010), all other studied systems 
beyond $\sim$70 kpc do not show any signs of any such variations.  
Overall, the outer halo population appears dissimilar from the dSph {\em stars}, while similarities with 
GCs {\em within} a dSph remain, as the example of the Fornax GCs shows, albeit at metallicities lower by 1 dex (Letarte et al. 2006). 
\section{Outer halo satellites -- the proper motions of Leo I and II}\label{sec:3}
As discussed above, a spectroscopic characterization of the remotest halo structures is feasible (e.g., Shetrone et al. 2009), yet very (exposure-) time consuming. 
Another way to tackle the question of a halo assembly is to study the dynamics of satellites; this point of view rather sheds light on the aspect of how much 
mass needs to be assembled onto the host system to end up with a halo as is observed and what the relative importance of individual, present-day satellites will be. 
For instance, the recent study of Watkins et al. (2010) has employed the kinematics of discrete tracers (field stars, GCs, dSphs) to estimate the total mass of the MW out to 300 kpc
as $(2.5\pm0.5)\times10^{12}$ M$_{\odot}$. This procedure is, however, sensitive to the in- or exclusion of kinematic outliers, such as the Leo~I and Hercules dSphs with their large distances and 
relatively high radial velocities (in the Galactic rest frame), which can alter the mass estimator by as much as $\sim$25\%. Knowledge of the full phase-space information, in particular the proper motions, of the 
tracers is required to ultimately construct a realistic mass model for the MW  and to assess the membership of any such system with the MW. 

At their large distances of 230--250 kpc, proper motion measurements for the Leo I and II dSphs are strictly not any ``easier'' than obtaining high S/N, high-resolution spectroscopy for their faint stars, as one is
chiefly dealing with sub-pixel motions (Anderson \& King 2000). Generally, at 100 kpc a transverse velocity of 100 km\,s$^{-1}$ corresponds to a proper motion of $\sim$0.2 mas\,yr$^{-1}$, or a mere 0.03 HST/WFPC2  pixels over a typical 
base line of 15 years. 

In fact, based on 14 years worth of archival HST data, anchored to a system of 17 extragalactic reference sources, we succeeded in  determining the proper motion of the remote Leo~II dSph (see L\'epine et al. 2011 for 
details and numbers).  The resulting, large space velocity in the Galactic rest frame ($v_{\rm GRF}=266\pm129$ km\,s$^{-1}$) is chiefly dominated by a large tangential component ($v_{\rm t}=265\pm129$ km\,s$^{-1}$), indicating that 
Leo~II is currently at apo- or pericenter, or on a highly eccentric orbit. The comparison with the local MW escape velocity (Fig.~4) indicates that this object is currently formally bound to the Galaxy at the 1$\sigma$-level; at the large distance of these tracers, our assessment is insensitive to the exact  choice of the MW potential. 
\begin{figure}
\begin{center}
\resizebox{0.7\columnwidth}{!}{
  \includegraphics{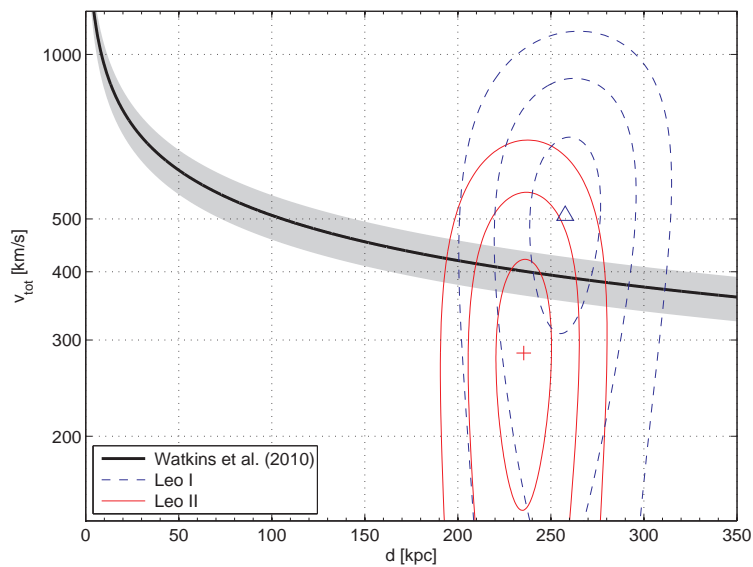} 
  }
\end{center}
\caption{Space velocity of Leo~II (L\'epine et al. 2011) and a preliminary measurement for Leo~I (L\'epine et al., in prep.) in comparison with the Galactic escape velocity (Watkins et al. 2010). Shown each are 1,2, and 3$\sigma$-contours.}
\label{fig:4}       
\end{figure}
On the other hand, the implied ``orbital'' period  amounts to 50 Gyr and its ``apocenter'' lies well outside 2 Mpc, which prohibits us to trace its exact orbital paths unless the entire Local Group's dynamic was accounted for. We conclude that Leo II has rather evolved in isolation (in concordance with its star formation history; Koch et al. 2007) and is now passing through the MW halo for the first time, as is seen also in M31 
(e.g., Majewski et al. 2007). 

While the  8\% fractional contribution of Leo II to the mass budget of the MW (Watkins et al. 2010) does not appear pivotal, the role of Leo~I  (at 27\%) is of prime importance.  
From a comparable HST data set we were able to measure a proper motion for the latter and the resulting, preliminary space velocity (Fig.~4) implies that Leo~I might not be bound to the MW, although this result is 
marginal at present (at 0.5$\sigma$) and needs to await consolidation from our careful analysis (L\'epine et al. in prep.). It is likely, that also this dSph has formed and evolved in isolation and is now approaching its first encounter with the 
(outer) halo of the Galaxy. Whether suchlike objects will actually shed enough stars to contribute  significantly to the halo field star population is, however, 
questionable and needs further orbital study. 
%
%
%
%

\end{document}